\documentclass[12pt,a4paper]{article}
\usepackage{graphicx}
\begin{document}

\begin{center}

{\large{\bf Theoretical interpretation of the experiments on
Parametric X-ray radiation in case of backward diffraction.}}
\vskip 4mm

{\normalsize  V.G.Baryshevsky, O.M.Lugovskaya

Institute for Nuclear Problems, Belarussian State University,
Minsk}

\bigskip

{\bf Abstract}

\end{center}

{\footnotesize

The spectral-angular and angular distributions of parametric
X-radiation for case of backward diffraction (particular case of
Bragg geometry scheme) is discussed. It is shown that in case of
Bragg geometry it is necessary to use dynamical approach for PXR
consideration. The comparison of the theory and experiment is
carried out.}

\section*{Introduction}

Since the theoretical prediction of Parametric X-radiation (PXR)
in crystals \cite{1}--\cite{4}  and its experimental observation
in 1985 \cite{5,6} a great number of experiments dedicated to
studying of PXR characteristics has been carried out. Most of
these experiments were performed in schemes of Laue geometry
(Fig.1a) and so called extremely asymmetric geometry, in which PXR
photons were emitted from the crystal through the lateral surface
of crystal plate at the right angle relative to the electron beam
velocity $\vec{v}$ (see Fig.1b).

\begin{figure}
\begin{center}
\includegraphics{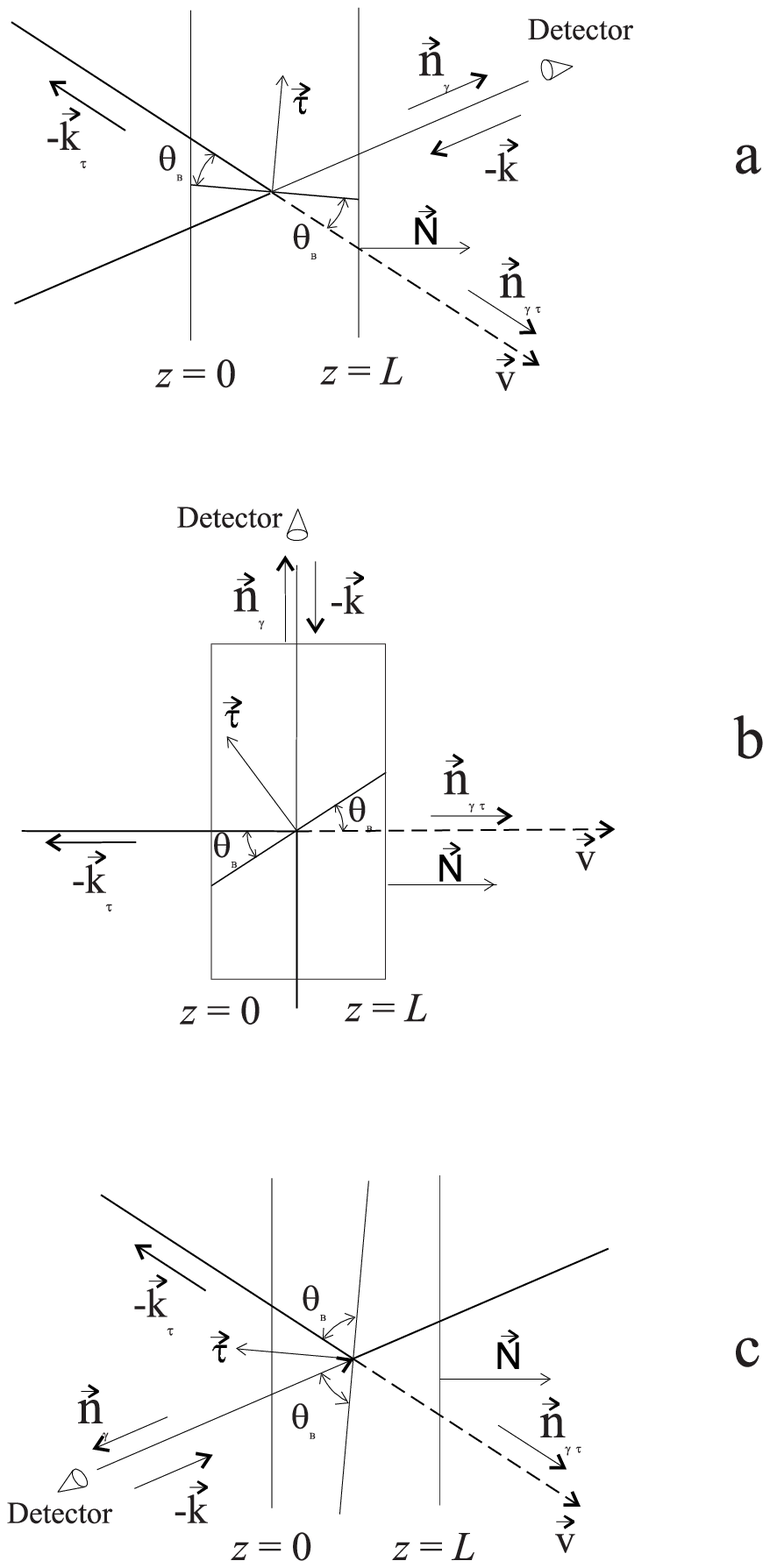}
\end{center}
\caption{\bf\footnotesize The schemes of different geometries of
PXR observation: a -- Laue, b -- Extremely asymmetrical, c --
Bragg.}
\end{figure}

In this case for theoretical interpretation of experimental data
it is sufficient to use simplified by the special way exact
theory, developed in \cite{7,8}. Specified simplification of the
theory is similar in some details to kinematical approximation
used by Ter-Mikaelyan for description of resonance radiation of
charged particle in medium with periodical dialectical
permittivity \cite{9}. Let us remind, that the typical property of
resonance radiation (and its main difference from PXR) is
dependence of emitted photons energy on the energy of charged
particles, while the PXR frequency is constant and determines only
by the crystal lattice period and direction of charged particle
propagation relative to crystallographic planes. However, the
attempts of using some simplified variants of the theory for
explanation of Bragg geometry experiments (see Fig.1c) appeared to
be unsuccessful. In this paper it is shown that for theoretical
description of experiments in Bragg scheme it is necessary to use
dynamical theory, developed in \cite{10,11}.

In that way, there's a series of experiments on PXR measurement in
scheme of Bragg geometry, which did not get any theoretical
interpretation. In the this work we present and discuss the
results of numerical calculations of spectral-angular and angular
distributions of PXR in backward geometry, which is a particular
case of Bragg geometry scheme, for the experimental parameters
corresponding to the Mainz microtrone \cite{12}. The experimental
data were kindly given to us by professor H.Backe with colleagues,
and as far as they are not published, we don't present them in
this paper. In addition, we performed calculations of angular
distributions for the experiment \cite{13}, which was also made in
the backward geometry.

\section {\bf General expressions for PXR spectral-angular  intensity in Bragg
 diffraction scheme}

The spectral- angular distribution of radiation, generated by the
charged particle at pass through the crystal plate into the
maximum at the angle $2\theta_B$  ($\theta_B$ -- angle between the
particle velocity vector $\vec{v}$ and planes corresponding to
vector $\vec{\tau}$) relative to the direction of its velocity in
scheme of Bragg diffraction is given by the following expression
\cite{11,14}:
\begin{equation}
\frac{d^{2}N_{s}}{d\omega d \vec{O} }=\frac{e^{2}Q^{2}\omega
}{4\pi ^{2}\hbar c^{3}}(\vec{e}_{\tau s}\vec{v})^{2} \left|
\sum_{\mu =1,2}\gamma _{\mu s}^{\tau }\left[ \frac{1}{\omega
-\vec{k}_{\tau } \vec{v}}-\frac{1}{\omega -\vec{k}_{\mu \tau
s}\vec{v}}\right] \left[ e^{\frac{i(\omega -\vec{k}_{\mu \tau
s}\vec{v})L}{c\gamma _{0}}}-1\right] \right| ^{2},
\end{equation}
$$\gamma_{1(2)s}^\tau = \frac{-\beta_1 C_s \chi_\tau}
{(2\varepsilon_{2(1)s}-\chi_0)-
 (2\varepsilon_{1(2)s}-\chi_0)\exp\left(i\frac{\omega}{c\gamma_0}(\varepsilon_{2(1)s}-
                                                        \varepsilon_{1(2)s})L\right)},$$
where $eQ$ -particle charge, $C_s=\vec{e}_s \vec{e}_{\tau s}$,
$\vec {e_{\tau 1}}\parallel\left[\vec{k}\vec{\tau} \right]$,
$\vec{e_{\tau 2}}\parallel \left[\vec{k} \vec{e_{\tau 1}}
\right]$-- the unit vectors of radiation polarization, $\vec{e}_s$
-- the unit polarization vector of incident wave, $\vec{k_{\mu
s}}= \vec{k}+\frac{\omega} {c\gamma_0}\varepsilon_{\mu s}
\vec{N}$, $\vec{N}$ - unit vector of the normal to the entrance
surface of a crystal plate, directed inside the crystal, $\chi_0$,
$\chi_\tau$, $\chi_{-\tau}$ -- Fourier-components of complex
crystal susceptibilities,
\begin{equation}
\varepsilon_{\mu s}=
\frac{1}{4}\left(-\alpha_B\beta_1+\chi_0(\beta_1+1)\pm
\sqrt{\left[-\alpha_B\beta_1+\chi_0(\beta_1-1) \right]^2
+4\beta_1\chi_\tau^s\chi_{-\tau}^s}\right),
\end{equation}
\begin{equation}
\alpha_B=\frac{2\vec{k}\vec{\tau}+\tau^2}{k^2},
\end{equation}
$\alpha_B$ is the Bragg-off parameter ($\alpha_B=0$ in case of
exact fulfillment of Bragg's condition)
$\beta_1=\gamma_0/\gamma_1$, $\gamma_0=\vec{n}_\gamma\vec{N}$,
$\vec{n}_\gamma=\frac{\vec{k}}{k}$,
$\vec{n}_{\gamma\tau}=\frac{\vec{k}+\vec{\tau}}{|\vec{k}+\vec{\tau}|}$,
$\gamma_1=\vec{n}_{\gamma\tau}\vec{N}$, $L_0$ - the thickness of
the crystal along the direction of a charged particle velocity
$L_0=L/\gamma_0$. The expression (1) has transparent physical
sense: in case of two-beam diffraction the crystal has two reflex
indexes $n_{\mu s}=\frac{k_{z\mu s}}{k_z}=1+\frac{\kappa_{\mu
s}}{k_z}$, axis $z$ is directed along vector $\vec{N}$,
$\kappa_{\mu s}=\frac{\omega}{c\gamma_0}\varepsilon_{\mu s}$.

Every item in (1) describes well-known radiation amplitude $A_{\mu
s}$ of photon arising as a result of charged particle movement
through the crystal target of $L$ thickness. Such as there are two
reflex indexes, that the total radiation density expressed through
the square of module of amplitudes sum, that is $\frac{d^2
N_s}{d\omega d\vec{O}}\sim |A_{1s}+A_{2s}|^2$.

Such as $\chi'_0<0$ though from the Vavilov-Cherenkov condition it
follows that only for a single root $(\mu=1)$ the real part of
refraction index $n'>1$.
 As a result the difference $\omega-\vec
k_{1\tau s}\vec v$ can turn into zero and the term of the
expression (1) comprising this difference in a denominator, begins
to grow proportionally $L$. At first sight it means that the term,
containing this difference (quasi-Cherenkov term), will give the
main contribution into the radiation when increasing the thickness
of the crystal along the particle velocity. However, in case of
Bragg diffraction there's a considerable distinction of physical
phenomenon, taking place in the crystal, from the case of Laue
diffraction, namely, in some area of frequencies and angles the
phenomenon of total reflection takes place. In this area of angles
and frequencies the wave vectors in crystal  lattice become
imaginary values at conditions of absorption absence. In the area
of total reflection, stipulated by the existence of the
heterogeneous wave in the crystal, it is necessary to take into
account of both dispersion branches during calculation of the
radiation intensity in Bragg diffraction scheme. Although the
structure of expression (1) is very simple, but in order to obtain
quantitative data it is necessary to performed correct
calculations on formula (1) with taking into account all terms of
it, since presence oscillating ones and its interference can
easily bring  to wrong results if some terms are neglected.

\section{\bf PXR in backward geometry }

    Recently there were conducted the experiments on observation of X-ray
radiation, generated by the charged particles in scheme of
backward diffraction \cite{13} - see the Fig 2.

\begin{figure}
\begin{center}
\includegraphics[width=11cm,height=8cm]{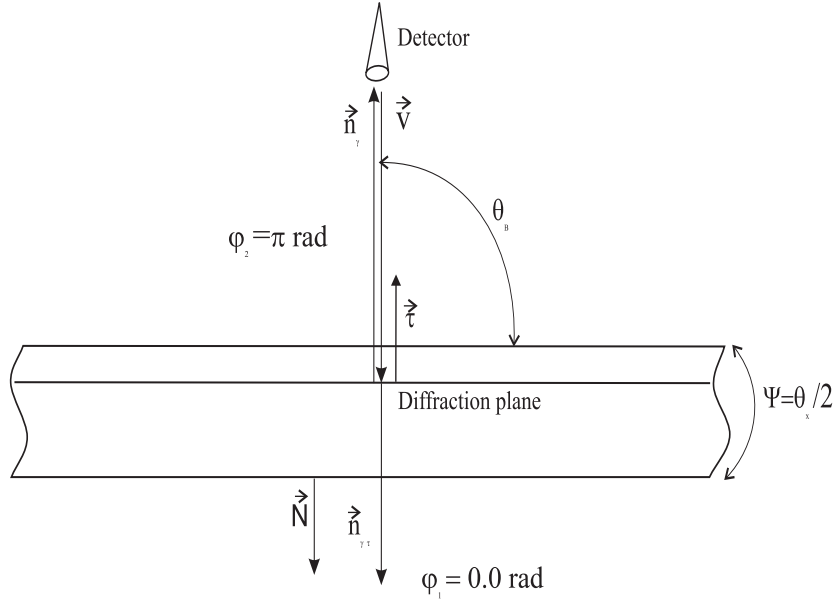}
\end{center}
\caption{\bf\footnotesize The scheme of backward diffraction
experiment, the angular distribution is taken as the function of
the tilt angle $\psi$.}
\end{figure}

 There were taken the angular dependence of radiation intensity as the function of
the tilt angle $\psi$ relatively the direction of charged
particles movement, for the tilt angle $\psi = 0$ this direction
coincides with the direction of charged particles movement, the
Bragg's angle $\theta_B=90^\circ$, the radiation is detected at
the angle $2\theta_B=180^\circ$ relative to this direction. At
rotation of the crystal at the angle $\psi$ the Bragg's angle
becomes equal to $\theta'_B=\theta_B+\psi$, the radiation angle --
$2\theta'_B=2\theta_B+\theta_X$, where $\theta_X=2\psi$.

The scheme of this kind (Fig.2) presents a keen interest, because
the theoretical description of the radiation intensity cannot be
given in scope of kinematical diffraction, such as existence of
the inhomogeneous wave brings to the possibility of realization of
the effect of the total reflection. Let us make a comparison of
numerical calculations results basing on the formula (1) with the
results, received in the experiment on the microtone in Mainz. The
measurements were taken in a silicon crystal plate thick $L=525$
 $\mu$m, the energy of electron beam $855$ MeV. The temperature of
the target was maintained at 120 K. There was studied the
radiation in reflexes (111), (333), (444), (555), (777), (888). At
the Fig.3 there are presented spectral-angular distributions of
PXR for the reflex (444), received basing on the formula (1) for
the tilt angles of the crystal 0,3 mrad, 2,5 mrad and 5 mrad.

\begin{figure}
\begin{center}
\includegraphics[width=11cm,height=8cm]{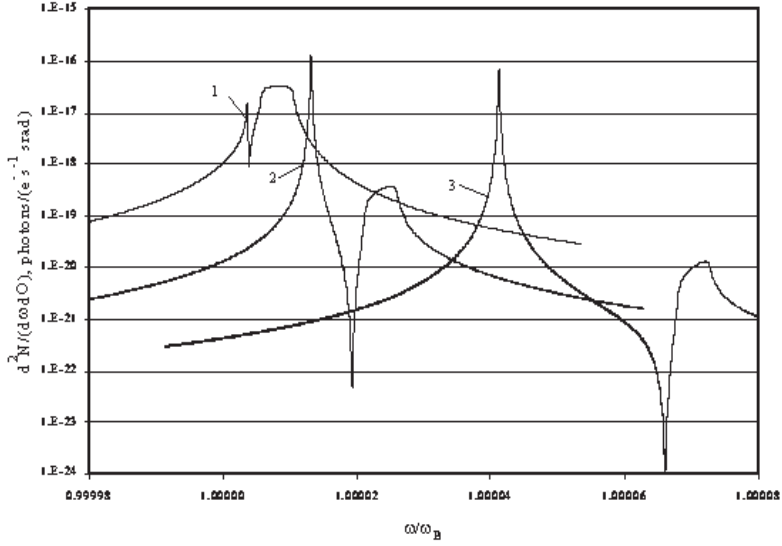}
\end{center}
\caption{\bf\footnotesize The spectra of the PXR radiation,
received for the tilt angles of the crystal: 1 - 0,3 mrad; 2 - 2,5
mrad; 3 - 5,0 mrad.}
\end{figure}

As far as the tilt angle gets greater there's a shift of the
spectrum towards increasing of the frequency. The narrow maximum
in the radiation spectrum corresponds at calculation by (1) to the
term, proportional to $\frac{1}{\omega-\vec {k}_{1\tau
\sigma}\vec{v}}$, which denominator can turn into $0$, and so this
peak can be interpreted as conditioned by quasi-Cerenkov radiation
mechanism.

With increasing of the tilt angle of the crystal $\psi$ (what
corresponds to the polar angle of the radiation $2\psi$) the
spectral-angular intensity of this maximum increases and at some
angle $\sim\vartheta_{ph}=\sqrt{\gamma^{-2}-\chi'_0}$ ($\gamma$ --
Lorentz factor of charged particle) becomes to exceed the
intensity of radiation maximum, intensity of which can be received
using simplified theory \cite{9} based on simple presentation of
ordinary transition radiation diffraction on crystallographic
planes, and maximum corresponds to area of maximal effectiveness
of the X-ray reflection on the crystal surfaces (the radiation,
emitted at angles and frequencies, for which Vavilov-Cherenkov
condition does not fulfill, however the coefficient of the X-ray
radiation reflection is maximum). It's necessary to note, that the
spectral width of the quasi-Cerenkov maximum is defined by the
expression $\frac{\Delta\omega}{\omega_B}\sim \frac{c}
{L_{eff}\omega_B\sin^2\theta_B}$ \cite{10}, here
$\omega_B=\frac{\pi c}{d\sin\theta_B}$ -- Bragg frequency, $d$ --
interplanar distance, $L_{eff}=min\left(L_0, L_{abs}\right)$, $
L_{abs}$ -- absorption length, while the width of maximum,
corresponds to area of total reflection
$\frac{\Delta\omega}{\omega_B}\sim
\frac{|\chi_\tau|}{\sin^2\theta_B}$.

For the reflex (444) the width of the quasi-Cherenkov maximum is
in order $4\times10^{-7}\omega_B$, what more than one degree
narrow of the width of peak in total reflection area, equal to
$\sim5\times10^{-6}\omega_B$. At the Fig.4 for the reflex (555)
there were demonstrated the spectral-angular distributions of
"forming" the sum radiation components (the radiation,
corresponding to different branches of dispersion curves
($\mu=1,2$). It's clear from the picture, that the quasi-Cherenkov
maximum belongs to the first dispersion branch $\mu=1$.

\begin{figure}
\begin{center}
\includegraphics[width=11cm, height=8cm]{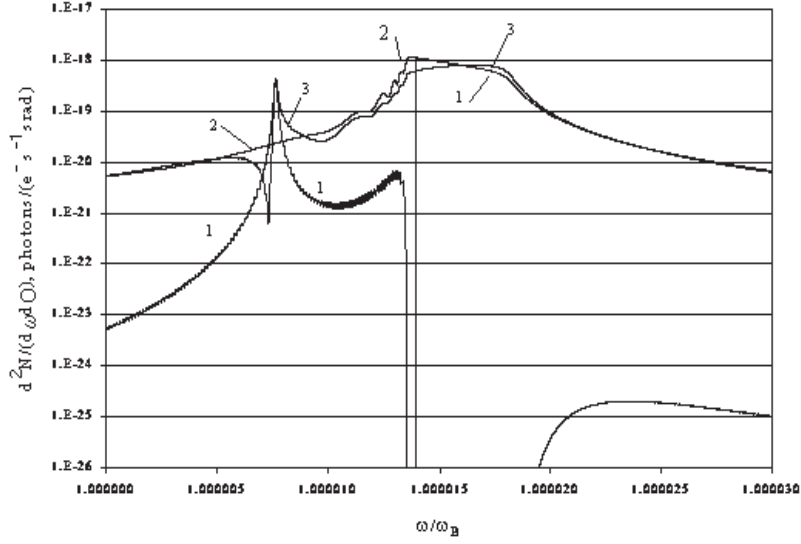}
\caption{\bf\footnotesize The spectrum of the radiation,
calculated by the formula (1) for 1 - $\mu=1$; 2 - $\mu=2$; 3 -
the sum radiation.}
\end{center}
\end{figure}

The distribution at the Fig.4 was received for the tilt angle of
the crystal $\psi=0,3$ mrad, at such angles the intensity of the
maximum, corresponding to area of total reflection is considerably
exceeding the intensity of the quasi-Cherenkov maximum. The
angular intensity at such angles is fully defined by total
reflection area, as the frequency's width of  this maximum is much
greater the width of quasi-Cherenkov maximum.

At the fig.5 there are presented the angular distributions of the
radiation as a function of the tilt angle for reflexes (111),
(333) and (444). The form of distributions coincides well with the
experimental curves, the value of the angular distribution in the
maximum for the reflex (111) is different from the experimental
one at $10\%$, for the reflex (333) at $18\%$, for (444) - at
$21\%$. So, this difference appears at small angles - less and
order of 0,3 mrad. The difference increases with increasing of the
energy of photons being emitted. Possible explanation of this
effect can be the influence of the multiple scattering, and
exactly, the additional contribution in the intensity of the
bremsstrahlung radiation, emitted at small angles.

\begin{figure}
\begin{center}
\includegraphics[width=11cm, height=8cm]{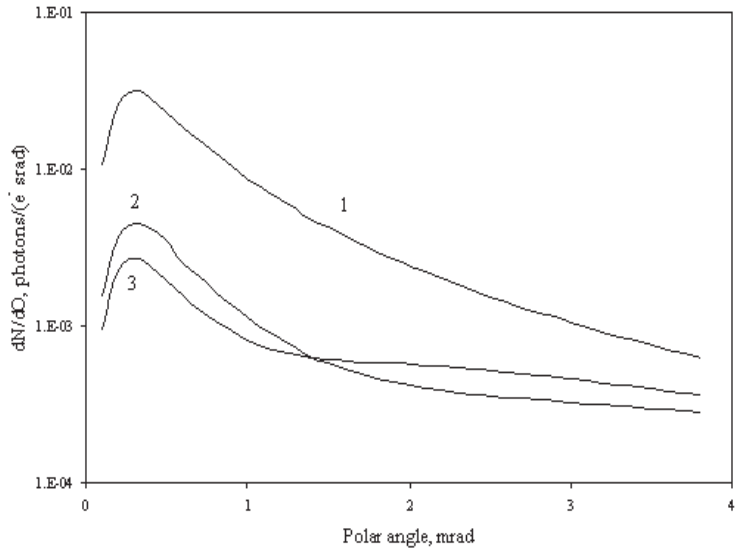}
\caption{\footnotesize The angular distribution in the reflexes: 1
- (111), the energy resolution of the detector
$\Delta\omega=358$eV, 2 - (333), $\Delta\omega=402$ eV, 3 - (444),
$\Delta\omega=400$ eV.}
\end{center}
\end{figure}

When decreasing the energy of charged particles, the contribution
of the bremsstrahlung radiation increases considerably. This is
explained by fact that coherent length of bremsstrahlung radiation
$L_{Br}=\sqrt{4c\over{\omega\overline{\theta^2_s}}}$,
$\overline{\theta^2_s}$ - root-mean-square angle of multiple
scattering (MS), becomes less then coherent length of PXR. In the
paper \cite{13} there is described an experiment also in backward
geometry, only at the energy of the electron beam one degree lower
($E_p=80,5$ and $86,5$ MeV). In this energy region coherent length
of bremsstrahlung radiation $L_{Br} \ll L$, and the account of
bremsstrahlung radiation contribution is very important. Our
calculations showed that the intensity of angular distribution in
the maximum, received by the integration of the expression (1)
over frequency in the range $\Delta\omega=10^{-3}\omega_B$,  is
two times lower the experimental value. However, it is necessary
to take into consideration the contribution of the bramsstrahlung
radiation into the sum radiation at such energies of electron beam
(estimations were made basing on the formulas from the paper
\cite{10}. In that way, it's possible to state that received
results for the angular intensity of the radiation agree well with
the experimental data.

\section*{Conclusion}
Conducted comparisons of theoretical and experimental distributions of PXR in
the backward geometry showed a good coincidence of results of the theory,
developed in \cite{11} and the experiment. It allows to state, that only the
approach of dynamical theory of the diffraction allows to interpret experiment.
The other simplified approaches, such as kinematical theory, approach in which
diffraction maximum is considered as diffracted on crystal planes transition
radiation (it is called in some works diffraction transition radiation (DTR))
and which does not take into account interference between different dispersion
branches \cite{15,16}, don't bring to successful interpretation of experimental
data.

\end{document}